\documentclass[12pt]{article}
\usepackage{amsmath,amsfonts}
 
\makeatletter\@addtoreset{equation}{section}\makeatother

\input epsf

\renewcommand{\title}[1]{\vbox{\center\LARGE{#1}}\vspace{5mm}}
\renewcommand{\author}[1]{\vbox{\center#1}\vspace{5mm}}
\newcommand{\address}[1]{\vbox{\center\em#1}}
\newcommand{\email}[1]{\vbox{\center\tt#1}\vspace{5mm}}

\begin{document}

\title{No-hair theorems for black holes \\
in the Abelian Higgs model}

\author{Juan Fern{\'a}ndez-Gracia, Bartomeu Fiol.}

\address{

Departament de F{\'\i}sica Fonamental i \\Institut de Ci{\`e}ncies del Cosmos, 

Universitat de Barcelona,\\

Mart{\'\i}\ i Franqu{\`e}s 1, 08193 Barcelona, Catalonia, Spain}

\email{jfernagr8@alumnes.ub.edu, bfiol@ub.edu}

\begin{abstract}

Motivated by the study of holographic superconductors, we generalize no-hair theorems for minimally coupled scalar fields charged under an Abelian gauge field, in arbitrary dimensions and with arbitrary horizon topology. We first present a straightforward generalization of no-hair theorems for neutral scalar hair. We then consider the existence of extremal black holes with scalar hair, and in the case of horizons with zero or positive curvature, provide a bound on the mass and charge of the scalar field that are necessary for the scalar hair to develop.

\end{abstract}

\section{Introduction and conclusions}
The gravity/gauge duality has established itself as a new paradigm to study quantum field theories in strongly interacting regimes. In the original examples of the duality, one has a good understanding of the theories on the two sides of the duality, and the approximations made are under control. However, none of these examples seems to be realized in Nature. One of the realms of physical phenomena where strongly coupled field theories ofter appear is that of condensed matter, and there has been plenty of work (some of it reviewed in \cite{Hartnoll:2009sz, Herzog:2009xv}) trying to apply holographic ideas  to condensed matter systems. These works deal with gravity backgrounds that do not necessarily correspond to limits of known M/string theory solutions, and its purpose is to present universal features that don't depend on details of a microscopic formulation of the holographic duality. Nevertheless, in more recent work some of these models have been embedded in string/M-theory \cite{Gubser:2009qm}  which makes possible a microscopic formulation of the holographic duality.

Among the possible applications of holographic techniques to condensed matter systems, quantum criticality has emerged as a potentially fruitful arena \cite{Sachdev:2008ba}. It refers to systems that in some region of the phase diagram have dynamics determined by the proximity of a quantum phase transition. The paucity of analytical tools to study these regions in phase space has raised the hope that gravity/gauge duality might provide instructive toy models, and more optimistically, perhaps even a precise dual for some real world system exhibiting quantum criticality.

Along these lines, there has been work trying to model the $T\neq 0$ quantum critical regime, by mapping it to an AdS background with a Reissner-Nordstr\"om black brane \cite{Hartnoll:2007ih}. There has been also work trying to model thermal (i.e. $T\neq 0$) phase transitions involving the quantum critical regime, by mapping them to the appearance of scalar \cite{Gubser:2008px, Hartnoll:2008vx, Hartnoll:2008kx} or vector \cite{Herzog:2009ci} hair in the dual black brane background. However, there has been much less work trying to use the gravity/gauge duality to model the quantum ($T=0$) phase transition itself (see however \cite{Gubser:2008wz}). This is one of our main motivations to explore the existence of possible extremal (i.e. $T=0$) solutions to the gravity systems considered in the literature. 

The kind of systems considered in these works consist of General Relativity  with negative cosmological constant coupled to either a non-Abelian gauge field, or to an Abelian gauge field and a minimally coupled charged scalar with an arbitrary potential. In the last case,  the resulting system of equations appears to be unsolvable in general, so much of the insight into it comes from numerical analysis  \cite{Gubser:2008px, Hartnoll:2008vx, Hartnoll:2008kx}. Numerical analysis is performed for particular choices of parameters (e.g. the charge of the scalar field) or models (e.g. specific choice of the scalar potential), and in some cases it is not conclusive, as in some limits the numerical behavior of the potential solutions is unstable. For these reasons, it is useful to complement these numerical analysis with analytic arguments. While finding analytic regular solutions is hard (see however \cite{Martinez:2004nb} for analytic solutions with hyperbolic horizons), it is possible to present arguments about non-existence of these solutions in certain cases. These no-hair theorems \cite{Bekenstein:1996pn, Mayo:1996mv, AyonBeato:1996kf} provide then constraints on the phase diagram of the putative dual theory. Most no-hair theorems in the literature apply to spherical horizons, but as we will see, they generalize immediately to flat horizons.

The main result of the present work is the proof of a couple of no-hair theorem for black holes in the Abelian Higgs model, in arbitrary dimension and for arbitrary horizon topology. These theorems help to elucidate the space of solutions of these models of holographic superconductivity as one varies the parameters of the model.

In section 2 we introduce the theory we will considering, General Relativity (in $d\geq 4$ dimensions) with cosmological constant coupled to an Abelian gauge field, a minimally coupled charged scalar field and an arbitrary scalar potential. We show that a particular combination of Maxwell and Einstein equations can be integrated, and the resulting relation is particularly simple for flat horizons, and for arbitrary horizons in four dimensions. Since we are interested in discussing properties of possible solutions near the horizon, we present our ansatz in coordinates manifestly adapted to the horizon, {\it i.e.} null Gaussian coordinates \cite{Reall:2002bh}, which render the discussion of regularity of the different fields quite transparent. Null Gaussian coordinates have become particularly useful in discussing extremal black holes and their possible near horizon geometries in a wide variety of 
contexts \cite{Kunduri:2007vf}.

In section 3 we generalize known no-hair theorems for neutral condensates to the case of flat horizons (black-branes). For flat horizons and the simplest potential $V(\psi)=m^2\psi^2$, the theorem that we present here covers most of the mass range where no hairy black branes are expected. There is, however, a range of masses for which no hairy black branes are expected \cite{Hartnoll:2008kx}, but we don't have an analytic argument ruling them out.

In section 4, we then turn to extremal ($T=0$) hairy black holes. We first find the possible near-horizon geometries compatible with our ansatz. We then show that in asymptotically Anti de Sitter spaces, the near horizon solution of flat or spherical horizons must have an $AdS_2$ factor in the metric, and it has vanishing scalar field. This result in turn implies that the effective mass of the scalar \cite{Gubser:2008px} at the horizon is bounded from below,
$$
m^2_{eff}(r_+)> 4|\Lambda|
$$
In particular, this rules out the existence of hairy extremal black branes with $m^2<0$ scalar hair. As a consequence, the endpoint of the instability discovered in \cite{Hartnoll:2008kx} is not an extremal hairy black brane.  

It is plausible that the bounds presented here can be improved by more elaborated arguments.
It would also be interesting to obtain analytic results delimiting the boundaries of existence/non-existence of hairy non-extremal black branes for the case of charged scalars.

{\bf Note added:} Since the appearance of the first version of this work, there have been a couple of papers \cite{Gubser:2009cg, Horowitz:2009ij} studying the $T=0$ limit of holographic superconductors. As these papers focus on solutions without finite area horizon, there is no direct overlap with the present work.

\section{The model and the ansatz}
We consider General Relativity in $d\geq 4$ dimensions with an arbitrary cosmological constant, plus a $U(1)$ gauge field and a minimally coupled charged scalar field with arbitrary scalar potential. 
\begin{equation}
{\cal L}=\sqrt{-g}\left( R-2\Lambda -\frac{1}{4}F^{\mu \nu}F_{\mu \nu}
-g^{\mu\nu}(\partial _\mu \psi-iqA_\mu \psi)(\partial _\nu \psi^*+iqA_\nu \psi^*)-V(|\psi|)\right)
\label{lagran}
\end{equation}
We require that $V(|\psi|)$ is such that $V(0)=0$. In the AdS/condensed matter literature, the following type of ansatz is ususally considered
\begin{equation}
\begin{array}{l}
ds^2=-f(r)dt^2+g^{-1}(r)dr^2+r^2h_{ij}(x)dx^idx^j \hspace{1cm}i,j=1,\dots, d-2\\
A_\mu(r)dx^\mu=\phi(r)dt\hspace{1cm}\psi=\psi(r)
\end{array}
\label{ansatz}
\end{equation}
where $h_{ij}$ is the metric of a $d-2$ Einstein manifold $\Sigma$ \cite{Birmingham:1998nr},
$$
R_{ij}(h)=(d-3)k h_{ij}(r)
$$
with $k=1,0,-1$. Solutions with $k=0,-1$ are known to have horizons only for $\Lambda <0$; the most relevant case for holographic applications to condensed matter is $k=0,\Lambda<0$. 

An example of solutions are the RN-AdS black holes \cite{Chamblin:1999tk}\footnote{These are still solutions for $\Lambda \geq 0$, but unless $k=1$, they don't correspond to black holes.}, 
\begin{equation}
f=g=\frac{r^2}{L^2}+k-\frac{A}{r^{d-3}}+\frac{B}{r^{2(d-3)}}
\label{rnsol}
\end{equation}
$$
\phi(r)=\phi_\infty-\frac{\rho}{r^{d-3}}\hspace{1cm}B=\frac{1}{2}\frac{d-3}{d-2}\rho^2
$$
where we introduced the AdS radius $L^2=-(d-2)(d-1)/2\Lambda$. For solutions with non-trivial $\psi$ it will be no longer true that $f=g$. Let us define $e^\chi=g/f$, which keeps track of the hairiness of the solution. Its equation of motion is given by a linear combination of Einstein equations,
$$
\frac{d-2}{2}\chi'+r\psi'^2+\frac{rq^2\phi^2\psi^2 e^\chi}{g^2}=0
$$
We presented the ansatz with a choice of coordinates that, while conventional, is not the most convenient one to discuss no-hair theorems, especially in the case of extremal black holes. The reason is that a key ingredient in some no-hair theorems is the requirement of regularity of curvature invariants at the horizon, which in turns dictates the allowed behavior of various combinations of metric components and matter fields \cite{Mayo:1996mv}. These conditions are more transparent if from the onset we switch to coordinates in which the metric is regular at the horizon, e.g. null Gaussian coordinates \cite{Reall:2002bh}. In the case at hand, we apply the following change of coordinates
\begin{equation}
dv=dt+\frac{dr}{(fg)^{1/2}}\hspace{.8cm}d\tilde r=\left(\frac{f}{g}\right)^{1/2}dr
\label{changeco}
\end{equation}
In these new coordinates the ansatz for the metric reads
$$
ds^2=-f(\tilde r)dv^2+2dvd\tilde r+h(\tilde r)h_{ij}(x)dx^idx^j 
$$
where $h(\tilde r)=r^2$. Einstein equations take the form
$$
\frac{d-2}{4}\frac{f'h'}{h}+\frac{d-2}{2}\frac{fh''}{h}+\frac{(d-5)(d-2)}{8}\frac{fh'^2}{h^2}
-\frac{(d-3)(d-2)k}{2h}+\Lambda  =  -\frac{1}{4}\phi'^2-\frac{V}{2}-\frac{q^2\phi^2\psi^2}{2f}
-\frac{1}{2}f \psi'^2
$$
$$
-\frac{d-2}{2}\frac{h''}{h}+\frac{d-2}{4}\left(\frac{h'}{h}\right)^2  =  \frac{q^2\phi^2\psi^2}{f^2}
+ \psi'^2
$$
$$
\frac{d-3}{2}\frac{f'h'+fh''}{h}+\frac{f''}{2}+\frac{(d-3)(d-6)}{8}\frac{fh'^2}{h^2}
-\frac{(d-3)(d-4)}{2h}k+\Lambda  =\frac{1}{4}\phi'^2-\frac{V}{2}+\frac{q^2\phi^2\psi^2}{2f}
-\frac{1}{2}f \psi'^2
$$
where the prime stands for $\tilde r$-derivative. Maxwell equation in the new coordinates is
$$
\phi''+\frac{d-2}{2}\frac{h'}{h}\phi'=\frac{2q^2\phi\psi^2}{f}
$$
while the equation of motion for the scalar field reads
$$
f\psi''+\left(f'+\frac{d-2}{2}\frac{h'}{h}f\right)\psi'+\frac{q^2\phi^2\psi}{f}=\frac{1}{2}\frac{dV}{d\psi}
$$
In the equations above we have already made use that Maxwell equations require the phase of the scalar field to be constant, and we take it to be real.

An important feature of this system of equations is that it admits a first integral. By taking a linear combination of the $ij$ minus the $v\tilde r$ components of Einstein equations, we arrive at an equation that does not depend on the scalar potential $V(\psi)$. If we further plug Maxwell equation in the resulting expression, we arrive at an equation that does not contain the scalar field, 
$$
\left(f'-\frac{h'}{h}f-\phi\phi'\right)'+\frac{d-2}{2}\frac{h'}{h}\left(f'-\frac{h'}{h}f-\phi\phi'\right)+\frac{d-3}{h}k=0
$$
This equation can be easily integrated in two particular cases. First, when $k=0$, it integrates to
\begin{equation}
\left(f'-\frac{h'}{h}f-\phi\phi'\right)h^{\frac{d-2}{2}}=c
\label{firstint}
\end{equation}
while for $d=4$ it integrates to
$$
\left(f'-\frac{h'}{h}f-\phi\phi'\right)h+k\tilde r=c
$$
For black hole solutions, the physical meaning of the constant $c$ will be elucidated after we discuss possible horizons.

\subsection{Horizons}
If the solution has $f(r_+)=0$ for some $r=r_+$, there is a horizon. The change of coordinates (\ref{changeco}) defining $\tilde r$ determines it up to an integration constant. For solutions with horizons, we fix this integration constant by demanding that $\tilde r=0$ at the outer horizon.
The horizon temperature is
$$
T=\frac{1}{4\pi}f'|_{\tilde r=0}
$$

We will restrict ourselves to solutions with non-zero area at the horizon; recalling that $h(\tilde r)=r^2$, this means that $h(0)\neq 0$. Furthermore, as argued for instance in \cite{Herzog:2009xv}, we require that $|A_\mu|^2=\phi^2/f$ is finite at the horizon. It then follows that $\phi$ itself vanishes at the horizon.
On the other hand, the manifest regularity of the metric at the horizon implies that the components of the stress energy tensor that appear in Einstein equations, and linear combinations thereof, are bounded at the horizon. In particular $\phi'$ must be regular at the horizon. Combining these last two facts, we conclude that $\phi\phi'$ vanishes at the horizon. A particular implication of this result is that  the constant $c$ appearing in (\ref{firstint}) is proportional to the temperature of the horizon, so for extremal solutions, 
$c=0$.\footnote{This equation, for $d=4$ and $k=0$ was derived independently in \cite{Gubser:2008gr}, where it was also noted that $c=0$ corresponds to extremal solutions.}

\section{A no-hair theorem for neutral scalars}
Before we address the possibility of extremal hairy black-branes, we present a no-hair theorem for another particular case, that of neutral ($q=0$) scalar hair. Since the argument relies on global properties of the putative solutions, and not just behavior near the horizon, it turns out that the coordinates in which we presented the ansatz (\ref{ansatz}) are better suited for this discussion.

In the context of holographic superconductors, it has been argued \cite{Hartnoll:2008kx} that at low enough $T$, for certain range of masses of the scalar field, a Reissner-Nordstr\"om black brane develops an instability due to a neutral scalar field, and numerical evidence was presented (for a particular value of the mass of the scalar) of the existence of black branes with neutral scalar hair. In this section we present a straightforward generalization of known no-hair theorems for neutral scalars \cite{Bekenstein:1995un} to arbitrary horizon topology and dimension. 

If we set $q=0$, the Maxwell equation can be readily integrated, yielding
\begin{equation}
\phi' e^{\chi/2} =\frac{(d-3)\rho}{r^{d-2}}
\label{truee}
\end{equation}
where we used the asymptotic behavior to fix an integration constant. We see that $\phi'$ has a definite sign (the sign of $\rho$). Since $\phi(r_+)=0$ and $\phi'$ has definite sign, it follows that $\phi $ has the same definite sign outside the horizon, i.e. $\phi\phi'>0$. Bringing this result into equation (\ref{firstint}) we conclude that in the case $k=0$, $f'>0$ outside the horizon. 

We will now show that for $\Lambda<0$, if $V(\psi)|_{r=r_+}\geq 0$ it is not possible to have neutral hair. 
First we present an extremely simple argument valid for flat horizons, $k=0$, and then we give a general argument valid for any $k$. The argument will be based on the conservation of the stress-energy tensor, which in the original coordinates takes the form

$$
0=\partial_r T_r^r+\frac{f'}{2f}\left(T^r_r-T_t^t\right)+\frac{d-2}{r}\left(T_r^r-T_i^i\right)
$$

\noindent
and the observation that if the scalar field is neutral, the stress -energy tensor of the scalar and vector fields are separately conserved. Consider the stress-energy tensor of the scalar field,
$$
T_r^r(\psi)=\frac{g\psi'^2}{2}-\frac{V}{2}
$$
at the horizon we have
$$
T_r^r(\psi) |_{r=r_+}=-\frac{V}{2} \leq 0
$$
Furthermore
$$
\partial_r T_r^r(\psi)=-\left(\frac{f'}{2f}+\frac{d-2}{r}\right)g\psi'^2 <0
$$
since we argued above that $f'>0$ outside the horizon. But these results are incompatible with the fact that near infinity
$$
T_r^r(\psi)\sim \frac{\lambda(d-1)}{2L^2}\frac{\psi_0^2}{r^{2\lambda}} \rightarrow 0^+
$$
This concludes the argument that it is not possible to have neutral hair if $V(\psi)_{r=r_+}\geq 0$. For the simple potential $V=m^2\psi^2$ this implies that there are no hairy solutions for $m^2\geq 0$. The  argument (and numerical evidence for $m^2L^2=-2$) in \cite{Hartnoll:2008kx} suggests that in 4d there are hairy solutions for $m^2L^2 < -3/2$. It would be interesting to extend the no-hair theorem presented here to the range $-3/2\leq m^2L^2 <0$.

For generic scalar potentials things can be more complicated. For instance, the analytic solution found recently in \cite{Zeng:2009fp} evades the theorem because $V(\psi) |_{r=r_+}< 0$. This is possible even for $m^2>0$, because terms in $V(\psi)$ with higher powers of $\psi$ become important near the horizon.

It is not much harder to extend the proof for arbitrary $k$. It is possible to present a proof using $T_r^r(\psi)$, but it is perhaps cleaner to use $e^{-\chi/2}T_r^r(\psi)$ as in \cite{Sudarsky:2002mk}. 
One has
$$
e^{\chi/2}\left(e^{-\chi/2}T_r^r(\psi)\right)'=-\left[-\frac{1}{4}e^\chi\phi'^2-\Lambda+\frac{(d-2)(d-3)k}{2r^2}+\frac{(d-2)(d-1)g}{2r^2}\right]\frac{r}{d-2}\psi'^2
$$
From the $rr$ Einstein equation at the horizon, it follows that
$$
-\frac{1}{4}e^{\chi_+}\phi'^2_+-\Lambda+\frac{(d-2)(d-3)k}{2r^2_+}+T_r^r(\psi_+)\geq 0
$$
Using this result and equation (\ref{truee}), it is immediate that for $k=0,-1$, 
$$
e^{\chi/2}\left(e^{-\chi/2}T_r^r(\psi)\right)'\leq \left[T_r^r(\psi_+)-\frac{(d-2)(d-1)g}{2r^2}\right]\frac{r}{d-2}\psi'^2
$$
while for $k=1$
$$
e^{\chi/2}\left(e^{-\chi/2}T_r^r(\psi)\right)'\leq \left[\hbox{max}\left(T_r^r(\psi_+),\Lambda\right)-\frac{(d-2)(d-1)g}{2r^2}\right]\frac{r}{d-2}\psi'^2
$$
(in fact, this last formula applies always, since for $k=0,-1$ it reduces to the previous one). In all cases, with $\Lambda\leq 0$, we see that if $T_r^r(\psi)|_{r=r_+}\leq 0$ it follows that $\left(e^{-\chi/2}T_r^r(\psi)\right)'<0$ but then it is impossible that $T_r^r(\psi)\rightarrow 0$ at infinity, so there is no such hairy solution. To recapitulate, we have proven the following

{\bf Theorem}: If $\Lambda<0$, $V(\psi)\geq 0$ and $q=0$, the ansatz (\ref{ansatz}) does not have regular black hole solutions with non-zero area.

\section{A no-hair theorem for extremal black branes}
In this section we consider the possible existence of extremal (i.e. $T=0$) black hole solutions for the Abelian Higgs model, within our ansatz. We first study possible near horizon geometries, finding that they can be of the form $AdS_2\times \Sigma$ or ${\mathbb R}^{1,1}\times \Sigma$, where $\Sigma $ is the $d-2$ Riemannian Einstein manifold that we introduced in our ansatz.  We then go on to consider possible full solutions, and argue that none of them has ${\mathbb R}^{1,1}\times \Sigma$ as its near horizon.

We then focus on the $k=0,1$ cases, for which we argue that the scalar field $\psi(\tilde r)$ vanishes in the near horizon geometry. Finally, for these cases, we derive a bound on the effective mass of the scalar field at the horizon. We conclude that for scalar fields with $(m,q)$ not satisfying this bound, regular extremal black holes with scalar hair are not possible.

\subsection{Near horizon geometry}
An important observation about solutions with extremal horizons is that they imply the existence of another solution to the system of equations: the near horizon geometry \cite{Reall:2002bh}. This new solution is obtained by taking the scaling limit
$$
\tilde r\rightarrow \tilde r\epsilon \hspace{1cm} v\rightarrow \frac{v}{\epsilon} 
\hspace{1cm} \epsilon \rightarrow 0
$$
Now let's assume that we have a solution with an extremal horizon, and consider its near horizon geometry. Define $f(\tilde r)=\tilde r^2F(\tilde r)$. Since $f(\tilde r)\geq 0$ at or outside the outer horizon, $F(\tilde r)\geq 0$. The near horizon metric is
$$
ds^2_{NH}=-\tilde r^2F(0)dv^2+2dvd\tilde r+h(0)h_{ij}dx^idx^j
$$
Recall that $F(0)\geq 0,h(0)>0$. The near horizon geometries are of the form $AdS_2\times \Sigma$ for $F(0)>0$ and ${\mathbb R}^{1,1}\times \Sigma$, for $F(0)=0$. As for the matter fields, since the Maxwell field strength is an antisymmetric 2-form, in the scaling limit it is reduced to a constant electric field. Finally, the scalar field, being a 0-form, is constant in the near horizon geometry. This can be also deduced from the equation of motion, since the near horizon geometry plus matter fields must solve them. To start with, the near horizon geometry must be a solution of Einstein equations. This implies
$$
-\frac{(d-3)(d-2)k}{2h(0)}+\Lambda  = -\frac{1}{4}\phi'^2 
- \frac{V}{2}-\frac{q^2\phi^2\psi^2}{2\tilde r^2F(0)}-\frac{1}{2}\tilde r^2F(0) \psi'^2
 $$
$$
0=\frac{q^2\phi^2\psi^2}{\tilde r^4F(0)^2}+\psi'^2
$$
$$
 F(0)-\frac{(d-3)(d-4)k}{2h(0)}+\Lambda =\frac{1}{4}\phi'^2 
- \frac{V}{2}+\frac{q^2\phi^2\psi^2}{2\tilde r^2F(0)}-\frac{1}{2}\tilde r^2F(0) \psi'^2
$$
Looking at the second equation, the vanishing of $G_{\tilde r\tilde r}(\tilde r)$ in the near horizon geometry has powerful implications, since it is equal to the sum of positive quantities, so each of them has to vanish separately. It follows that for any value of $d,k$, the scalar field $\psi$ is constant in the near horizon geometry, and from its equation of motion in the near horizon geometry, its value must be an extremum of the potential $V(\psi)$. Also $q\phi\psi$ vanishes everywhere in the near horizon geometry. For charged scalars, that means that either $\psi(\tilde r)$ or $\phi(\tilde r)$ vanish everywhere in the near horizon geometry. Maxwell equation in the near horizon geometry does not add any new information.

We have then learned that both $\phi'(\tilde r)$ and $\psi(\tilde r)$ are constant in the near horizon geometry, 
\begin{eqnarray}
\frac{1}{2}\phi'(\tilde r)^2& = & F(0)+\frac{(d-3)k}{h(0)} \\
V(\psi(\tilde r)) & = & -F(0)+\frac{(d-3)^2}{h(0)}k-2\Lambda \nonumber 
\label{nhconstants}
\end{eqnarray}
and at least $\phi$ or $\psi$ vanish everywhere for this solution. Let's discuss these cases in turns.

Consider first the case $\phi=0$. If $F(0)>0$, then $k=-1$. The only extremal black hole that we are aware having this near horizon geometry is actually a vacuum solution, obtained by setting $A=-\frac{2}{d-1}\left(\frac{d-3}{d-1}L^2\right)^{(d-3)/2}, B=0$ in (\ref{rnsol}). We are not aware of any non-vacuum extremal solution in the literature whose near horizon is given by this geometry: the family of black holes with scalar hair of \cite{Martinez:2004nb} admits a solution with an extremal horizon, but in that case there is a curvature singularity at the horizon. Still in the $\phi=0$ case, if $F(0)=0$, then $k=0$, and $V(\psi)=-2\Lambda$. For $\Lambda <0$, recalling that in the near horizon geometry $\psi(\tilde r)$ takes a constant value given by an extremum of the potential, we can immediately rule out the appearance of this near horizon geometry for some simple potentials. For instance, if $V(\psi)=m^2\psi^2$, $V$ vanishes at its only extremum, so this choice of potential doesn't allow for extremal black branes with this near horizon. Also, if $V(\psi)=m^2\psi^2+\lambda\psi^4$, the potential doesn't have an strictly positive extremum for $m^2<0$, so again with this choice of potential there are no black branes with this near horizon geometry.

On the other hand, if $\phi$ is not identically zero in the NH geometry, it follows that $\psi=0$. It is easy to see that for $\Lambda<0$, it must be now that $F(0)>0$, since if $F(0)=0$, then $k=1$ and $\Lambda>0$. One can easily check that the resulting solutions are the near horizon geometries of the extremal Reissner-Nordstr\"om solutions (\ref{rnsol}).

\subsection{A global result}
In the classification of possible near horizon geometries we found that they were of the form $AdS_2 \times \Sigma$ or ${\mathbb R}^{1,1}\times \Sigma$, when $F(0)>0$ or $F(0)=0$ respectively. We would like to argue now that only the first case can appear as the near horizon geometry of a full extremal black hole solution. Assume that $F(0)=0$. This can only happen when $\phi=k=0$ in the near horizon geometry, so $V(\psi)=-2\Lambda$ and in particular $\psi\neq 0$ in the near horizon geometry. Now recall the first integral (\ref{firstint}) for the $k=0$ case. For extremal solutions $c=0$, and the expression is rather simple
$$
f'-\frac{h'}{h}f=\phi\phi'
$$
Using that $h(\tilde r)$ is regular and non-zero at the horizon, we learn that in this case
$$
\frac{\phi^2}{f}\rightarrow 2
$$
at the horizon. Requiring now that $T_{\tilde r\tilde r}$ is finite at the horizon, we deduce from this last result that $\psi$ vanishes at the horizon, in contradiction with $V(\psi)=-2\Lambda$. We then conclude that all extremal black holes of the Abelian Higgs model have near horizon geometries of the form $AdS_2 \times \Sigma$. 

We are now ready to present the main result of this section. We have already seen that all extremal black holes within our ansatz present a near horizon of the form $AdS_2\times \Sigma$. For $k=0,1$ this  near horizon geometry supports a non-vanishing constant electric field while the scalar field vanishes. For $k=-1$ we haven't ruled out the possibility of a non-vanishing value for the scalar fied in the near horizon, which would complicate the analysis. Therefore, in what follows we focus on the $k=0,1$ cases.

When $k=0,1$ we just saw that the scalar field vanishes in the near horizon geometry. This implies that in a potential full solution, the scalar field must vanish at the horizon. More than that, since 

$$
\frac{\phi^2}{2f}\rightarrow 1+\frac{(d-3)k}{F(0)h(0)}
$$

\noindent
tends to a finite non-zero value, finiteness of $T_{\tilde r \tilde r}$ at the horizon requires then that $\psi(\tilde r)$ vanishes at least linearly at the horizon.  Let's now consider the equation of motion for the scalar field in the full geometry. Every other field appearing in this equation has its leading behavior near the horizon fixed by the near horizon geometry,

$$
f(\tilde r)=\tilde r^2 F(0)+\dots \; , \hspace{.5cm}\phi(\tilde r)=\phi'(0)\tilde r+\dots \;,
\hspace{.5cm} h(\tilde r)=h(0)+\dots
$$

Therefore, in a perturbative expansion near the horizon, the leading behavior of the scalar field 
$\psi(\tilde r)$ is given by solving its equation of motion in the near horizon geometry. When we do so, we find two solutions, only one of them with a chance of vanishing at the horizon

$$
\psi(\tilde r)=\psi_+\tilde r^{\alpha_+}+\psi_-\tilde r^{\alpha_-} \; \;, \hspace{1.5cm} \alpha_{\pm}=
\frac{-1\pm\sqrt{1+\frac{4m_{eff}^2}{F(0)}}}{2}
$$

We argued that regularity of $T_{\tilde r\tilde r}$ demands that $\psi(\tilde r)$ in the full solution vanishes at least linearly in $\tilde r$ near the horizon. Imposing this requirement on the solution with $\alpha_+$, yields the necessary condition
$$
m^2_{eff}(r_+)=m^2-2q^2\left( 1+\frac{(d-3)k}{F(0)h(0)}\right)\geq 2F(0)=\frac{2(d-3)^2}{h(0)}k-4\Lambda
$$
or more succintly
$$
m^2_{eff}(r_+)\geq 4|\Lambda|
$$

We have proven the following:

{\bf Theorem}: If $\Lambda<0$ and $m^2-2q^2<4|\Lambda|$, the ansatz (\ref{ansatz}) does not contain
$k=0,1$ regular extremal black hole solutions with non-zero area and non-trivial scalar hair.

This bound rules out the existence of hairy extremal black branes with $m^2\leq 0$. In \cite{Hartnoll:2008kx} it was found that  an extremal RN black brane coupled to a scalar field with mass square below the $AdS_2$ BF bound develops an instability (see also \cite{Denef:2009tp}); no claims were made about the endpoint of the instability. Our result implies that the endpoint of that instability is not  a hairy extremal black hole.

Although the bound presented here rules out the existence of regular extremal black holes in the case  of interest for holographic superconductors, it leaves open the possibility of  solutions above this bound. It would be interesting to completely rule out their existence or alternatively, to find them explicitely.

\section{Acknowledgements} 
We would like to thank Roberto Emparan, Sean Hartnoll, Chris Herzog and Gary Horowitz for discussions and comments on the draft. We are especially grateful to Sean Hartnoll for sharing unpublished notes on \cite{Hartnoll:2008kx}. We would also like to thank the anonymous referee who suggested using null Gaussian coordinates to tighten the analysis of no-hair theorems. The research of J. F.-G. was supported by a Spanish MEPSYD fellowship for undergraduate students. The research of B.F is supported by a Ram\'on y Cajal fellowship, by grants FPA2007-66665C02-02 and  DURSI 2005-SGR-00082, and by the CPAN CSD2007-00042 project of the Consolider-Ingenio 2010 program.

\end{document}